\documentclass[aps,prl,reprint,superscriptaddress]{revtex4-1}

\usepackage[colorlinks,linkcolor=blue, urlcolor=blue, anchorcolor=blue, citecolor=blue]{hyperref}

\usepackage{graphicx}
\usepackage{dcolumn}
\usepackage{bm}

\draft 

\begin{document}


\title{Pattern Transitions in a Soft Cylindrical Shell} 


\author{Yifan Yang}
\affiliation{Institute of Mechanics and Computational Engineering, Department of Aeronautics and Astronautics, Fudan University, Shanghai 200433, P.R. China}
\author{Hui-Hui Dai}
\affiliation{Department of Mathematics, City University of Hong Kong, Hong Kong, P.R. China}

\author{Fan Xu}
\email[Corresponding author.\\]{fanxu@fudan.edu.cn}
\affiliation{Institute of Mechanics and Computational Engineering, Department of Aeronautics and Astronautics, Fudan University, Shanghai 200433, P.R. China}

\author{Michel Potier-Ferry}
\affiliation{Universit\'e de Lorraine, CNRS, Arts et M\'etiers ParisTech, LEM3, F-57000 Metz, France}


\date{\today}

\begin{abstract}
Instability patterns of rolling up a sleeve appear more intricate than the ones of walking over a rug on floor, both characterized as uniaxially compressed soft-film/stiff-substrate systems. This can be explained by curvature effects. To investigate pattern transitions on a curved surface, we study a soft shell sliding on a rigid cylinder by experiments, computations and theoretical analyses. We reveal a novel post-buckling phenomenon involving multiple successive bifurcations: smooth-wrinkle-ridge-sagging transitions. The shell initially buckles into periodic axisymmetric wrinkles at the threshold and then a wrinkle-to-ridge transition occurs upon further axial compression. When the load increases to the third bifurcation, the amplitude of the ridge reaches its limit and the symmetry is broken with the ridge sagging into a recumbent fold. It is identified that hysteresis loops and the Maxwell equal-energy conditions are associated with the co-existence of wrinkle/ridge or ridge/sagging patterns. Such a bifurcation scenario is inherently general and independent of material constitutive models.
\end{abstract}


\maketitle

\begin{figure*}[!htbp]
\includegraphics[width=16cm]{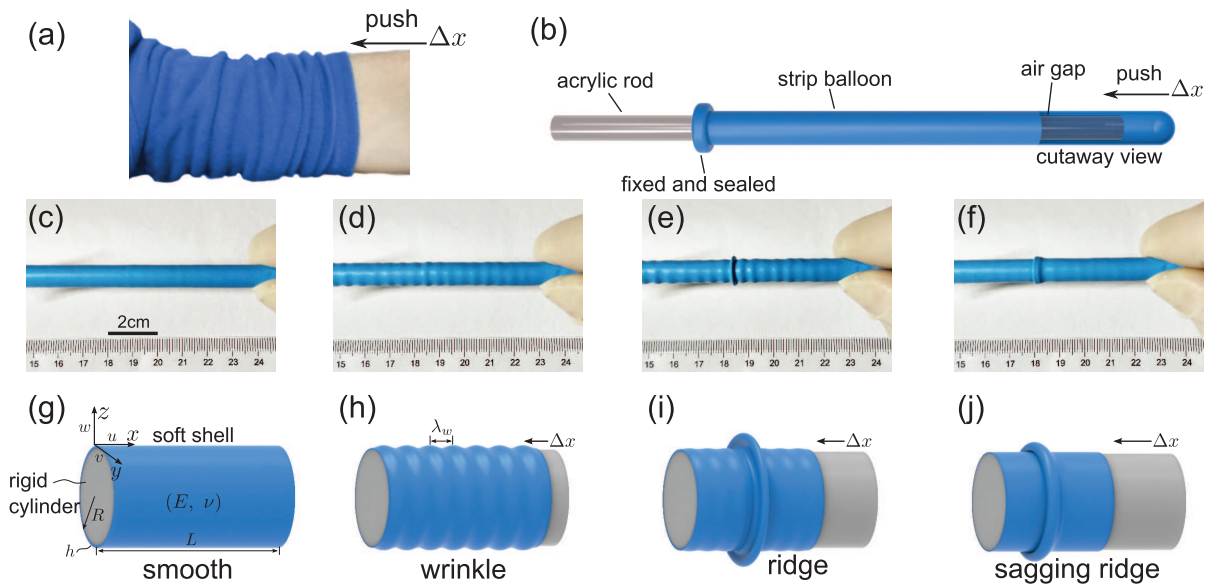}
\caption{Experiments of soft shells sliding on rigid cylinders. (a) Patterns upon rolling up a sleeve. (b) An air-inflated latex balloon of thickness $h=0.2$ mm with one end fixed on an acrylic rod of radius $R=3$ mm. With increasing compression from the right side, four distinguished states are observed: (c) initial undeformed configuration, (d) wrinkle, (e) ridge, (f) sagging ridge. (g)-(j) are representative sketches corresponding to (c)-(f), respectively.}
\label{fig:mod}
\end{figure*}

Surface topography of film/substrate systems has raised considerable interest not only because of the wide existence in nature but also for various functional applications. Abundant examples in living creatures across length scales include hierarchical wrinkling of skins \cite{Efimenko2005}, morphological buckling of fruits and vegetables \cite{Yin2008}, folding of growing tubular organs \cite{Ciarletta2014}, and differential growth of bacterial biofilms \cite{Zhang2016}. Applications through harnessing surface instabilities range from micro/nano morphological patterning control \cite{Stoop2015a}, reversible optical writing/erasure of functional interface \cite{Zong2016}, moisture-responsive wrinkling design with tunable dynamics \cite{Zeng2017}, to defect localization in elastic surface crystals \cite{Jimenez2016}. Most efforts have been devoted to film/substrate systems typically having a stiff thin layer attached on a soft substrate. Periodic solutions are generally obtained both in planar \cite{Cai2011} and curved \cite{Xu2016, Xu2017} geometries, even though excess stresses may break symmetry and lead to localized \cite{Jin2015} or disordered \cite{Li2011} instability patterns. One may wonder what would happen in the opposite way: a soft thin layer lying on a stiff foundation? One might be familiar with the motion and deformation of compressing a paper on the table from both ends: the paper globally bends to form a single fold. This simple but important buckling phenomenon can be used to understand how dislocations can facilitate the relative motion of two crystalline planes \cite{Nabarro1982}, and to characterize upheaval deformations of subsea pipelines \cite{Karampour2013}. Vella \textit{et al.} \cite{Vella2009} applied a linear combination of the tension and kinetic energy to investigate the inertial motion of a ruck in a rug. Kolinski \textit{et al.} \cite{Kolinski2009} demonstrated how the ruck becomes asymmetric and moves by rolling on an inclined plane. Both studies are constrained to planar geometry. What would happen on a curved surface? Ubiquitous cases can be found in the process of rolling up a sleeve \cite{Stoop2015} as shown in Fig. \ref{fig:mod}(a) and the twinkling of an eye (eyelid ``slides" on the eyeball) \cite{Zhu2013}. Up to date, the underlying mechanism for large deformations of curved soft-shell/stiff-substrate systems remains unclear.

We first design a demonstrative experiment to explore this nonlinear response. A strip latex balloon, representing the soft elastic shell, is wrapped around a rigid acrylic cylinder, as shown in Fig. \ref{fig:mod}(b). A small amount of air is sealed up in the balloon with one end fixed. The narrow air gap between the shell and the cylinder can reduce the friction at the interface and allows the shell to smoothly slide upon laterally compressing its free end. Yet the air gap should be kept sufficiently small so that the rigid cylinder can support the shell as long as the shell is compressed. The radius of the shell and the cylinder, therefore, can be treated as equal. The shell in Fig. \ref{fig:mod}(c) is initially cylindrical with smooth surface. When subject to axial compression, it loses stability and buckles into periodic axisymmetric wrinkles at the critical threshold as illustrated in Fig. \ref{fig:mod}(d). Upon further compression, we observe a wrinkle-to-ridge transition, where the wrinkles evolve into a localized response with one single ridge growing, at the expense of the intermediate undulations that are becoming almost smooth, as shown in Fig. \ref{fig:mod}(e). With increasing load, the amplitude of the ridge finally reaches its limit and the symmetry is broken with the ridge sagging into a recumbent fold, as depicted in Fig. \ref{fig:mod}(f).

\begin{figure}[!htbp]
\includegraphics[width=8.6cm]{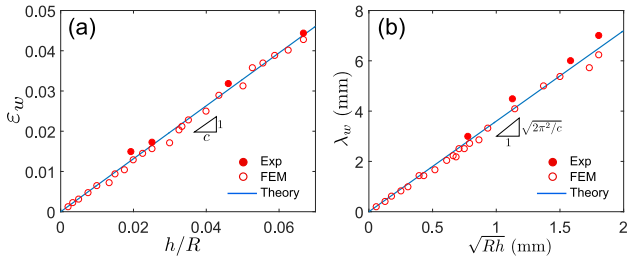}
\caption{Comparison of theoretical, numerical and experimental results at the critical wrinkling strain $\varepsilon_w$: (a) critical axial strain $\varepsilon_{w}$ as a linear function of dimensionless curvature $h/R$, (b) wrinkling wavelength $\lambda_{w}$ as a linear function of $\sqrt{Rh}$.}
\label{fig:koi}
\end{figure}

To understand the underlying mechanism, we consider a soft shell of thickness $h$, length $L$, elastic modulus $E$ and Poisson's ratio $\nu$, sliding on a rigid cylinder, as shown in Fig. \ref{fig:mod}(g). The radius of cylinder is denoted as $R$. Since the deformations in experiments (Fig. \ref{fig:mod}(c)-(f)) are axisymmetric (circumferential displacement $v=0$), the axial displacement $u$ and radial one $w$ can be assumed to be a function of $x$ for simplicity. The overall compressive strain can be defined as $\varepsilon=\Delta x/L$. The total potential energy of the system $U_{tot}$ can be written as the sum of the membrane energy $U_{m}$, bending energy $U_{b}$ and external work $U_{f}$ \cite{Sup}:
\begin{equation}
U_{tot} = U_{m} + U_{b} + U_{f}.
\label{eq:energy}
\end{equation}
Here the contact is ignored in the model. As an ansatz, we consider the following forms for the displacements in the wrinkling range: $u=A \sin(px/R)$, $w=C \cos(px/R)$, where $p$ represents the wave number along the axial direction, and $A$ and $B$ refer to the amplitudes of waves. Within the Donnell-Mushtari-Vlassov (DMV) framework \cite{Koiter1945, Yamaki1984, Koiter2009}, the minimization of the system (\ref{eq:energy}) yields the critical compressive strain and its wrinkling wavelength for an axially compressed shell \cite{Sup}:
\begin{equation}\label{eq:koi}
\varepsilon_{w} = \frac{h}{cR},\quad \lambda_{w} = \pi \sqrt{\frac{2Rh}{c}},
\end{equation}
where $c=\sqrt{3 (1-\nu^2)}$. The latex shell is nearly incompressible with Poisson's ratio $\nu\approx 0.48$. Substituting it into Eq. (\ref{eq:koi}), one obtains the onset of wrinkling state:
\begin{equation}
\varepsilon_{w} \approx 0.658h/R,\quad \lambda_{w} \approx 3.604 \sqrt{Rh}.
\label{eq:ana}
\end{equation}
This solution indicates the curvature effect, \textit{e.g.,} when $R \rightarrow \infty$, the wavelength $\lambda_w \rightarrow \infty$, which explains why a global buckling mode, namely a ruck (a single fold) \cite{Vella2009, Kolinski2009}, rather than wrinkles occurs when a flat paper is biaxially compressed on the table \cite{Sup}. To investigate the post-buckling morphological evolution, we apply finite element method (FEM) by accounting for various geometric parameters and constitutive laws \cite{Sup}. A four-node bilinear axisymmetric quadrilateral element (CAX4R) with reduced integration is adopted and mesh convergence was carefully examined. Between the soft shell and the rigid cylinder, Hertzian contact that suggests there is non-adhesive behavior and only normal pressure on the contact surface without friction, is taken into account. In fact, for instabilities that are extremely localized, \textit{e.g.,} ridge in Fig. \ref{fig:mod}(i) and sagging ridge in Fig. \ref{fig:mod}(j), there may exist a local transfer of strain energy from one part of the model to the neighboring parts, and global resolution methods may encounter convergence difficulties. This class of problems can be solved by implementing a pseudo-dynamic algorithm \cite{Sup}. Let the shell be laterally compressed from free stress state to an excess axial strain. We find that the critical values of FEM simulations and theoretical predictions in Eq. (\ref{eq:ana}) agree well with experiments (see Fig. \ref{fig:koi}), which implies that the Hertzian contact has no apparent influences on the periodic wrinkling state.

\begin{figure}[!htbp]
\includegraphics[width=8.5cm]{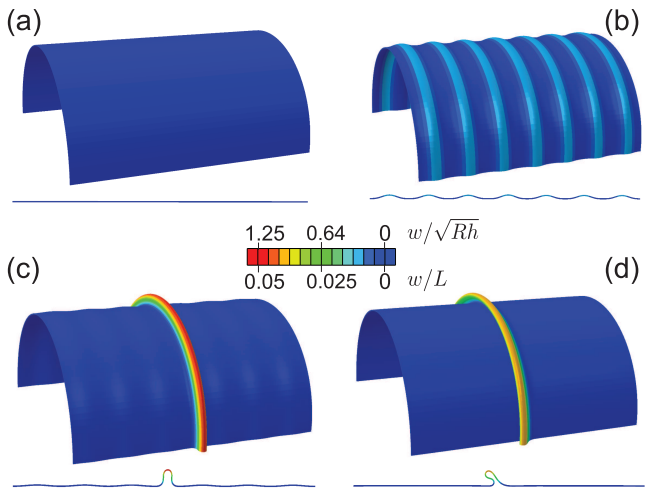}
\caption{FEM simulations on pattern evolutions of a soft-shell/rigid-cylinder system under axial compression: (a) initial undeformed state, (b) wrinkle, (c) ridge, and (d) sagging ridge. In simulations, we took $R=10$ mm and $h=0.1$ mm.}
\label{fig:nsp}
\end{figure}

\begin{figure}[!htbp]
\includegraphics[width=9cm]{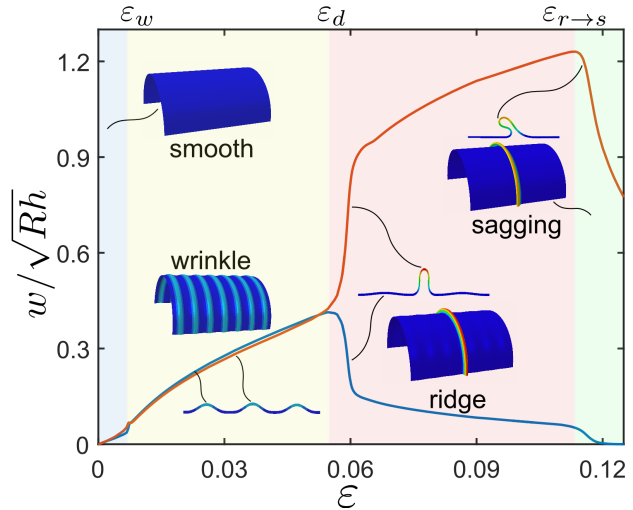}
\caption{Bifurcation diagram of dimensionless deflections $w/\sqrt{Rh}$ of the ridge peak (red curve) and its neighboring wrinkle peak (blue curve) as a function of overall compressive strain $\varepsilon$ ($R=10$ mm and $h=0.1$ mm). Three successive bifurcations and four phase regions are distinguished: smooth (blue), wrinkle (yellow), ridge (red), and sagging ridge (green).}
\label{fig:load}
\end{figure}

Figure \ref{fig:nsp} illustrates four representative patterns with increasing loading, consistent with the experimental observations in Fig. \ref{fig:mod}. The bifurcation portrait of dimensionless radial displacement $w/\sqrt{Rh}$ of the ridge peak and its neighboring wrinkle peak is plotted in Fig. \ref{fig:load}, based on the neo-Hookean constitutive law \cite{Holzapfel2000} (comparison with other hyperelastic material models is provided in \cite{Sup}). Under axial compression, the shell initially buckles into a periodic axisymmetric wrinkling mode at the critical load $\varepsilon_{w}=0.0066$. Note that the wrinkle mode is generally stable with a supercritical bifurcation (without hysteresis response) \cite{Xu2016}. Upon further compression, we observe a wrinkle-to-ridge transition, where the wrinkles evolve into a localized response, with one single ridge growing at the expense of the intermediate wrinkles that tend to be almost smooth simultaneously. The ridge peak and its neighboring wrinkle peak depart at $\varepsilon_{d}=0.055$, and experience a transition interval during which the ridge is rapidly developed until $\varepsilon_{w \rightarrow r}=0.069$. When the load increases to the third bifurcation at $\varepsilon_{r \rightarrow s}=0.118$, the amplitude of the ridge reaches its limit and the symmetry is broken with the ridge sagging into a recumbent fold. Note that the sagging process can release the strain energy of the vicinity and make the surface become almost smooth.

\begin{figure}[!htbp]
\includegraphics[width=8.5cm]{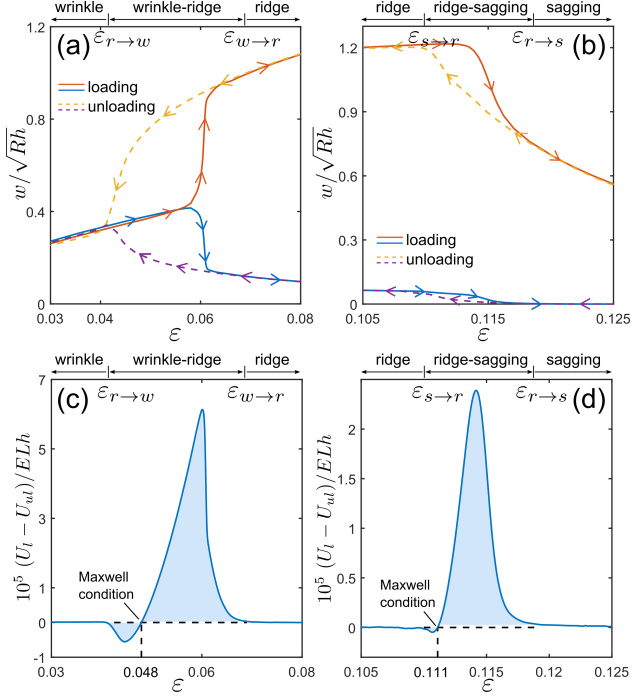}
\caption{Comparison between loading and unloading stages ($R=10$ mm and $h=0.1$ mm). The first line shows the deflections $w$ of the ridge/sagging peak and its neighboring wrinkle peak as a function of overall compressive strain $\varepsilon$ within hysteresis loops: (a) wrinkle/ridge transition, (b) ridge/sagging transition. (c) and (d) demonstrate the elastic energy differences between the loading and unloading states, $U_l - U_{ul}$, as a function of compressive strain $\varepsilon$, corresponding to (a) and (b), respectively. In the range $\varepsilon_{r \rightarrow w}<\varepsilon<\varepsilon_{w \rightarrow r}$ ($U_l - U_{ul}=U_w-U_r$) and $\varepsilon_{s \rightarrow r}<\varepsilon<\varepsilon_{r \rightarrow s}$ ($U_l - U_{ul}=U_r-U_s$), the Maxwell equal-energy conditions are defined by the points when $U_l - U_{ul}=0$.}
\label{fig:maxw}
\end{figure}

To further explore the physics and mechanics of wrinkle-ridge-sagging transitions, we look into a complete loading/unloading cycle. The deflections $w$ of the ridge peak with its neighboring wrinkle peak are plotted in Figs. \ref{fig:maxw}(a) and \ref{fig:maxw}(b), where the progression along the curves in the loading and unloading segments of the path history is indicated by the direction of the arrows. During unloading, the recumbent ridge reverts to the upright ridge again at $\varepsilon_{s \rightarrow r}=0.109$, smaller than $\varepsilon_{r \rightarrow s}=0.118$ in the loading stage. With further unloading, in the range beyond $0.109 <\varepsilon<0.118$, the solution is again reversible and consistent with the loading stage. The ridge returns to the wrinkles at a smaller overall strain $\varepsilon_{r \rightarrow w}=0.041$ compared with loading stage ($\varepsilon_{w \rightarrow r}=0.069$). In the range $\varepsilon<\varepsilon_{r \rightarrow w}$ until the shell comes back to the undeformed configuration, the wrinkle solution is again reversible and consistent with the loading stage, with undulations of the shell vanishing simultaneously. Viewing the whole unloading process, two hysteresis loops are found within the wrinkle/ridge and ridge/sagging transition stages, while the unbuckled/wrinkle transformation shows no hysteresis behavior. More attention is, therefore, paid to the hysteresis cycles. For wrinkle/ridge transitions, the shell is first compressed monotonically from $\varepsilon=0$ to an overall strain $\varepsilon=0.08$ which is larger than the wrinkle-to-ridge transition strain $\varepsilon_{w \rightarrow r}$, but smaller than the ridge-to-sagging strain $\varepsilon_{r \rightarrow s}$, and then unloaded monotonically back to $\varepsilon=0$. Note that there is significant hysteresis in the range $\varepsilon_{r \rightarrow w}<\varepsilon<\varepsilon_{w \rightarrow r}$ (see Fig. \ref{fig:maxw}(a)), with wrinkles prevailing during loading and ridges dominating during unloading. At $\varepsilon_{r \rightarrow w}$, the ridge reverts back to the periodic wrinkle state. The wrinkle/ridge transitions, therefore, seem to be characterized by two critical values, $\varepsilon_{r \rightarrow w}$ and $\varepsilon_{w \rightarrow r}$. Similarly, for ridge/sagging transitions as shown in Fig. \ref{fig:maxw}(b), the loading/unloading cycle is set within $\varepsilon=0.105\sim0.125$. The ridge/sagging transition is, therefore, governed by two critical values, $\varepsilon_{r \rightarrow s}$ and $\varepsilon_{s \rightarrow r}$. Yet it is still difficult to judge what happened during wrinkle-ridge-sagging coexistent stages. Further insights will be obtained from examining the differences in the elastic energy in the shell between the loading and unloading states, $U_l-U_{ul}$, associated with hysteresis cycles.

Pattern transitions in the post-buckling regime generally imply the change of energy states. We first focus on wrinkle/ridge transformations. When the energies of the wrinkle state $U_{w}$ and the ridge state $U_{r}$ share the same value of an overall axial compression, it is defined as the Maxwell strain $\varepsilon_{Max}^{wr}$ \cite{Hunt2000}. In the wrinkle/ridge transition region, the ridge state has a higher energy than the wrinkle state ($U_r - U_w > 0$), when the overall strain is lower than the Maxwell strain ($\varepsilon < \varepsilon_{Max}^{wr}$), while the ridge state has a lower energy ($U_r - U_w < 0$), when $\varepsilon > \varepsilon_{Max}^{wr}$. We further examine the elastic energy difference in the shell between the loading and unloading stages, associated with a hysteresis loop, which is also the difference between the energy in the wrinkle state and the ridge state, $U_w-U_r$, in the range $\varepsilon_{r \rightarrow w}<\varepsilon<\varepsilon_{w \rightarrow r}$. This difference is reported as a function of overall compressive strain $\varepsilon$ in Fig. \ref{fig:maxw}(c). Outside the Maxwell region (blue parts), the energies coincide $U_l-U_{ul}=0$ since only one stable solution exists. In this case, the Maxwell strain, $\varepsilon_{Max}^{wr} = 0.048$, is much closer to the ridge-to-wrinkle transition, $\varepsilon_{r \rightarrow w}=0.041$, than to the wrinkle-to-ridge transition, $\varepsilon_{w \rightarrow r}=0.069$. When $\varepsilon_{r \rightarrow w}<\varepsilon<\varepsilon_{w \rightarrow r}$, a significant difference in energies between the wrinkle and ridge states exists. At the Maxwell condition with $\varepsilon=\varepsilon_{Max}^{wr}$, the wrinkle state and the ridge state hold the same energy and, in principle, can coexist. Under such conditions, if $\varepsilon$ is then increased above $\varepsilon_{Max}^{wr}$, the ridge advances replacing wrinkles, while if $\varepsilon$ is reduced below $\varepsilon_{Max}^{wr}$, the wrinkles advance engulfing the ridge. A similar mechanism is observed for ridge/sagging transitions in Fig. \ref{fig:maxw}(d), where the Maxwell strain $\varepsilon_{Max}^{rs}=0.111$, associated with that the energy of the ridge state, $U_{r}$, equals to the sagging ridge state $U_{s}$, \textit{i.e.,} $U_r-U_s=0$. In \cite{Stoop2015}, Stoop \textit{et al.} did not observe the smooth-wrinkle-ridge transformation but one single fold, and predict that the symmetry-breaking bifurcation of a single fold occurs at $\Delta x \approx 1.88 \sqrt{Rh}$, which is equivalent to $\varepsilon=0.094$ for this case. The obvious periodic wrinkling undulations, however, cannot be ignored, especially when the dimensionless curvature $h/R$ becomes relatively large \cite{Sup}. The existence of small undulations outside the ridge region can fairly increase the global energy and make the sagging strain ($\varepsilon_{Max}^{rs}=0.111$) a little larger than the consideration of a single fold ($\varepsilon=0.094$). Note that the axisymmetric responses could be somehow surprising by comparing with earlier experiments achieved in 1960s \cite{Almroth1964, Horton1965}, where an internal solid mandrel was introduced to artificially stabilize deflections in the shell and all those experiments revealed non-axisymmetric patterns. For explaining this difference, we have performed full 3D simulations \cite{Sup}, showing a strong influence of the internal gap between the shell and the core. With a gap in the order of the shell thickness, non-axisymmetric diamond-like patterns were obtained in consistent with the experiments in \cite{Horton1965}, while the deformation remains constantly axisymmetric in the absence of a gap \cite{Sup}.

In summary, we have uncovered the post-buckling evolution and morphological pattern transitions of a soft cylindrical shell sliding on a rigid core. A remarkable finding, both revealed experimentally and numerically, is a novel multiple-bifurcation phenomenon, \textit{i.e.,} smooth-wrinkle-ridge-sagging transitions. The critical wrinkling strain is found linearly proportional to $\sim h/R$, and quantitatively consistent with analytical predictions. It seems that the hard contact somewhere at the interface between the soft shell and the rigid core has rather limited influence on the initial bifurcation. In the limit case of zero curvature ($R\rightarrow\infty$), the wavelength turns out to be infinite and thus a single fold emerges instead of periodic undulations, which coincides with the motion of a ruck in a rug \cite{Vella2009, Kolinski2009}. Upon further compression, a wrinkle-to-ridge transition is observed. Within the coexistence stage of wrinkles and ridge, the wrinkle mode is stable even if the compressive strain has been above the Maxwell strain $\varepsilon_{Max}^{wr}$, which implies there must exist a substantial energy barrier (subsequent bifurcation) between the wrinkle and ridge states in the range $\varepsilon_{r \rightarrow w}<\varepsilon<\varepsilon_{w \rightarrow r}$. In contrast, once the ridge has formed, it persists under unloading to the lower transition strain $\varepsilon_{r \rightarrow w}$. When the load increases to the third bifurcation, the amplitude of the ridge reaches its limit, and then the ridge sags into a recumbent fold due to symmetry breaking. In the ridge/sagging transitions, the Maxwell equal-energy condition exists as well and the mechanism is similar to the wrinkle/ridge transformation, both of which appear to be subcritical bifurcations, while the unbuckled/wrinkle transition is supercritical and reversible. The whole nonlinear evolution is numerically proven to be inherently general and independent of constitutive laws \cite{Sup}. Understanding the nonlinear morphological transitions of soft shells can promise an interesting fabrication strategy for multi-functional surfaces.

This work is supported by the National Natural Science Foundation of China (Grants No. 11602058 and No. 11772094), Shanghai Education Development Foundation and Shanghai Municipal Education Commission (Shanghai Chenguang Program, Grant No. 16CG01), and a GRF grant (Project No. CityU 11302417) from the Research Grants Council of Hong Kong SAR, China. F.X. is grateful for support from the Research Fellowship at Department of Mathematics, CityU. F.X. and M.P.F. acknowledge the financial support from the French National Research Agency ANR (LabEx DAMAS, Grant No. ANR-11-LABX-0008-01).



%
%

%


\bibliography{prl}

\begin{thebibliography}{27}%
\makeatletter
\providecommand \@ifxundefined [1]{%
 \@ifx{#1\undefined}
}%
\providecommand \@ifnum [1]{%
 \ifnum #1\expandafter \@firstoftwo
 \else \expandafter \@secondoftwo
 \fi
}%
\providecommand \@ifx [1]{%
 \ifx #1\expandafter \@firstoftwo
 \else \expandafter \@secondoftwo
 \fi
}%
\providecommand \natexlab [1]{#1}%
\providecommand \enquote  [1]{``#1''}%
\providecommand \bibnamefont  [1]{#1}%
\providecommand \bibfnamefont [1]{#1}%
\providecommand \citenamefont [1]{#1}%
\providecommand \href@noop [0]{\@secondoftwo}%
\providecommand \href [0]{\begingroup \@sanitize@url \@href}%
\providecommand \@href[1]{\@@startlink{#1}\@@href}%
\providecommand \@@href[1]{\endgroup#1\@@endlink}%
\providecommand \@sanitize@url [0]{\catcode `\\12\catcode `\$12\catcode
  `\&12\catcode `\#12\catcode `\^12\catcode `\_12\catcode `\%12\relax}%
\providecommand \@@startlink[1]{}%
\providecommand \@@endlink[0]{}%
\providecommand \url  [0]{\begingroup\@sanitize@url \@url }%
\providecommand \@url [1]{\endgroup\@href {#1}{\urlprefix }}%
\providecommand \urlprefix  [0]{URL }%
\providecommand \Eprint [0]{\href }%
\providecommand \doibase [0]{http://dx.doi.org/}%
\providecommand \selectlanguage [0]{\@gobble}%
\providecommand \bibinfo  [0]{\@secondoftwo}%
\providecommand \bibfield  [0]{\@secondoftwo}%
\providecommand \translation [1]{[#1]}%
\providecommand \BibitemOpen [0]{}%
\providecommand \bibitemStop [0]{}%
\providecommand \bibitemNoStop [0]{.\EOS\space}%
\providecommand \EOS [0]{\spacefactor3000\relax}%
\providecommand \BibitemShut  [1]{\csname bibitem#1\endcsname}%
\let\auto@bib@innerbib\@empty
\bibitem [{\citenamefont {Efimenko}\ \emph {et~al.}(2005)\citenamefont
  {Efimenko}, \citenamefont {Rackaitis}, \citenamefont {Manias}, \citenamefont
  {Vaziri}, \citenamefont {Mahadevan},\ and\ \citenamefont
  {Genzer}}]{Efimenko2005}%
  \BibitemOpen
  \bibfield  {author} {\bibinfo {author} {\bibfnamefont {K.}~\bibnamefont
  {Efimenko}}, \bibinfo {author} {\bibfnamefont {M.}~\bibnamefont {Rackaitis}},
  \bibinfo {author} {\bibfnamefont {E.}~\bibnamefont {Manias}}, \bibinfo
  {author} {\bibfnamefont {A.}~\bibnamefont {Vaziri}}, \bibinfo {author}
  {\bibfnamefont {L.}~\bibnamefont {Mahadevan}}, \ and\ \bibinfo {author}
  {\bibfnamefont {J.}~\bibnamefont {Genzer}},\ }\href {\doibase
  10.1038/nmat1342} {\bibfield  {journal} {\bibinfo  {journal} {Nat. Mater.}\
  }\textbf {\bibinfo {volume} {4}},\ \bibinfo {pages} {293} (\bibinfo {year}
  {2005})}\BibitemShut {NoStop}%
\bibitem [{\citenamefont {Yin}\ \emph {et~al.}(2008)\citenamefont {Yin},
  \citenamefont {Cao}, \citenamefont {Li}, \citenamefont {Sheinman},\ and\
  \citenamefont {Chen}}]{Yin2008}%
  \BibitemOpen
  \bibfield  {author} {\bibinfo {author} {\bibfnamefont {J.}~\bibnamefont
  {Yin}}, \bibinfo {author} {\bibfnamefont {Z.}~\bibnamefont {Cao}}, \bibinfo
  {author} {\bibfnamefont {C.}~\bibnamefont {Li}}, \bibinfo {author}
  {\bibfnamefont {I.}~\bibnamefont {Sheinman}}, \ and\ \bibinfo {author}
  {\bibfnamefont {X.}~\bibnamefont {Chen}},\ }\href {\doibase
  10.1073/pnas.0810443105} {\bibfield  {journal} {\bibinfo  {journal} {Proc.
  Natl. Acad. Sci. USA}\ }\textbf {\bibinfo {volume} {105}},\ \bibinfo {pages}
  {19132} (\bibinfo {year} {2008})}\BibitemShut {NoStop}%
\bibitem [{\citenamefont {Ciarletta}\ \emph {et~al.}(2014)\citenamefont
  {Ciarletta}, \citenamefont {Balbi},\ and\ \citenamefont
  {Kuhl}}]{Ciarletta2014}%
  \BibitemOpen
  \bibfield  {author} {\bibinfo {author} {\bibfnamefont {P.}~\bibnamefont
  {Ciarletta}}, \bibinfo {author} {\bibfnamefont {V.}~\bibnamefont {Balbi}}, \
  and\ \bibinfo {author} {\bibfnamefont {E.}~\bibnamefont {Kuhl}},\ }\href
  {\doibase 10.1103/physrevlett.113.248101} {\bibfield  {journal} {\bibinfo
  {journal} {Phys. Rev. Lett.}\ }\textbf {\bibinfo {volume} {113}},\ \bibinfo
  {pages} {248101} (\bibinfo {year} {2014})}\BibitemShut {NoStop}%
\bibitem [{\citenamefont {Zhang}\ \emph {et~al.}(2016)\citenamefont {Zhang},
  \citenamefont {Li}, \citenamefont {Huang}, \citenamefont {Ni},\ and\
  \citenamefont {Feng}}]{Zhang2016}%
  \BibitemOpen
  \bibfield  {author} {\bibinfo {author} {\bibfnamefont {C.}~\bibnamefont
  {Zhang}}, \bibinfo {author} {\bibfnamefont {B.}~\bibnamefont {Li}}, \bibinfo
  {author} {\bibfnamefont {X.}~\bibnamefont {Huang}}, \bibinfo {author}
  {\bibfnamefont {Y.}~\bibnamefont {Ni}}, \ and\ \bibinfo {author}
  {\bibfnamefont {X.~Q.}\ \bibnamefont {Feng}},\ }\href {\doibase
  10.1063/1.4963780} {\bibfield  {journal} {\bibinfo  {journal} {Appl. Phys.
  Lett.}\ }\textbf {\bibinfo {volume} {109}},\ \bibinfo {pages} {143701}
  (\bibinfo {year} {2016})}\BibitemShut {NoStop}%
\bibitem [{\citenamefont {Stoop}\ \emph {et~al.}(2015)\citenamefont {Stoop},
  \citenamefont {Lagrange}, \citenamefont {Terwagne}, \citenamefont {Reis},\
  and\ \citenamefont {Dunkel}}]{Stoop2015a}%
  \BibitemOpen
  \bibfield  {author} {\bibinfo {author} {\bibfnamefont {N.}~\bibnamefont
  {Stoop}}, \bibinfo {author} {\bibfnamefont {R.}~\bibnamefont {Lagrange}},
  \bibinfo {author} {\bibfnamefont {D.}~\bibnamefont {Terwagne}}, \bibinfo
  {author} {\bibfnamefont {P.~M.}\ \bibnamefont {Reis}}, \ and\ \bibinfo
  {author} {\bibfnamefont {J.}~\bibnamefont {Dunkel}},\ }\href {\doibase
  10.1038/nmat4202} {\bibfield  {journal} {\bibinfo  {journal} {Nat. Mater.}\
  }\textbf {\bibinfo {volume} {14}},\ \bibinfo {pages} {337} (\bibinfo {year}
  {2015})}\BibitemShut {NoStop}%
\bibitem [{\citenamefont {Zong}\ \emph {et~al.}(2016)\citenamefont {Zong},
  \citenamefont {Zhao}, \citenamefont {Ji}, \citenamefont {Xue}, \citenamefont
  {Xie}, \citenamefont {Wang}, \citenamefont {Cao}, \citenamefont {Jiang},\
  and\ \citenamefont {Lu}}]{Zong2016}%
  \BibitemOpen
  \bibfield  {author} {\bibinfo {author} {\bibfnamefont {C.}~\bibnamefont
  {Zong}}, \bibinfo {author} {\bibfnamefont {Y.}~\bibnamefont {Zhao}}, \bibinfo
  {author} {\bibfnamefont {H.}~\bibnamefont {Ji}}, \bibinfo {author}
  {\bibfnamefont {X.~H.}\ \bibnamefont {Xue}}, \bibinfo {author} {\bibfnamefont
  {J.}~\bibnamefont {Xie}}, \bibinfo {author} {\bibfnamefont {J.}~\bibnamefont
  {Wang}}, \bibinfo {author} {\bibfnamefont {Y.}~\bibnamefont {Cao}}, \bibinfo
  {author} {\bibfnamefont {S.}~\bibnamefont {Jiang}}, \ and\ \bibinfo {author}
  {\bibfnamefont {C.}~\bibnamefont {Lu}},\ }\href {\doibase
  10.1002/anie.201510796} {\bibfield  {journal} {\bibinfo  {journal} {Angew.
  Chem. Int. Ed.}\ }\textbf {\bibinfo {volume} {55}},\ \bibinfo {pages} {3931}
  (\bibinfo {year} {2016})}\BibitemShut {NoStop}%
\bibitem [{\citenamefont {Zeng}\ \emph {et~al.}(2017)\citenamefont {Zeng},
  \citenamefont {Li}, \citenamefont {Freire}, \citenamefont {Garbellotto},
  \citenamefont {Huang}, \citenamefont {Smith}, \citenamefont {Hu},
  \citenamefont {Tait}, \citenamefont {Bian}, \citenamefont {Zheng},
  \citenamefont {Zhang},\ and\ \citenamefont {Sun}}]{Zeng2017}%
  \BibitemOpen
  \bibfield  {author} {\bibinfo {author} {\bibfnamefont {S.}~\bibnamefont
  {Zeng}}, \bibinfo {author} {\bibfnamefont {R.}~\bibnamefont {Li}}, \bibinfo
  {author} {\bibfnamefont {S.~G.}\ \bibnamefont {Freire}}, \bibinfo {author}
  {\bibfnamefont {V.~M.}\ \bibnamefont {Garbellotto}}, \bibinfo {author}
  {\bibfnamefont {E.~Y.}\ \bibnamefont {Huang}}, \bibinfo {author}
  {\bibfnamefont {A.~T.}\ \bibnamefont {Smith}}, \bibinfo {author}
  {\bibfnamefont {C.}~\bibnamefont {Hu}}, \bibinfo {author} {\bibfnamefont
  {W.~R.}\ \bibnamefont {Tait}}, \bibinfo {author} {\bibfnamefont
  {Z.}~\bibnamefont {Bian}}, \bibinfo {author} {\bibfnamefont {G.}~\bibnamefont
  {Zheng}}, \bibinfo {author} {\bibfnamefont {D.}~\bibnamefont {Zhang}}, \ and\
  \bibinfo {author} {\bibfnamefont {L.}~\bibnamefont {Sun}},\ }\href {\doibase
  10.1002/adma.201700828} {\bibfield  {journal} {\bibinfo  {journal} {Adv.
  Mater.}\ }\textbf {\bibinfo {volume} {29}},\ \bibinfo {pages} {1700828}
  (\bibinfo {year} {2017})}\BibitemShut {NoStop}%
\bibitem [{\citenamefont {Jim\'enez}\ \emph {et~al.}(2016)\citenamefont
  {Jim\'enez}, \citenamefont {Stoop}, \citenamefont {Lagrange}, \citenamefont
  {Dunkel},\ and\ \citenamefont {Reis}}]{Jimenez2016}%
  \BibitemOpen
  \bibfield  {author} {\bibinfo {author} {\bibfnamefont {F.~L.}\ \bibnamefont
  {Jim\'enez}}, \bibinfo {author} {\bibfnamefont {N.}~\bibnamefont {Stoop}},
  \bibinfo {author} {\bibfnamefont {R.}~\bibnamefont {Lagrange}}, \bibinfo
  {author} {\bibfnamefont {J.}~\bibnamefont {Dunkel}}, \ and\ \bibinfo {author}
  {\bibfnamefont {P.~M.}\ \bibnamefont {Reis}},\ }\href {\doibase
  10.1103/physrevlett.116.104301} {\bibfield  {journal} {\bibinfo  {journal}
  {Phys. Rev. Lett.}\ }\textbf {\bibinfo {volume} {116}},\ \bibinfo {pages}
  {104301} (\bibinfo {year} {2016})}\BibitemShut {NoStop}%
\bibitem [{\citenamefont {Cai}\ \emph {et~al.}(2011)\citenamefont {Cai},
  \citenamefont {Breid}, \citenamefont {Crosby}, \citenamefont {Suo},\ and\
  \citenamefont {Hutchinson}}]{Cai2011}%
  \BibitemOpen
  \bibfield  {author} {\bibinfo {author} {\bibfnamefont {S.}~\bibnamefont
  {Cai}}, \bibinfo {author} {\bibfnamefont {D.}~\bibnamefont {Breid}}, \bibinfo
  {author} {\bibfnamefont {A.~J.}\ \bibnamefont {Crosby}}, \bibinfo {author}
  {\bibfnamefont {Z.}~\bibnamefont {Suo}}, \ and\ \bibinfo {author}
  {\bibfnamefont {J.~W.}\ \bibnamefont {Hutchinson}},\ }\href {\doibase
  10.1016/j.jmps.2011.02.001} {\bibfield  {journal} {\bibinfo  {journal} {J.
  Mech. Phys. Solids}\ }\textbf {\bibinfo {volume} {59}},\ \bibinfo {pages}
  {1094} (\bibinfo {year} {2011})}\BibitemShut {NoStop}%
\bibitem [{\citenamefont {Xu}\ and\ \citenamefont
  {Potier-Ferry}(2016)}]{Xu2016}%
  \BibitemOpen
  \bibfield  {author} {\bibinfo {author} {\bibfnamefont {F.}~\bibnamefont
  {Xu}}\ and\ \bibinfo {author} {\bibfnamefont {M.}~\bibnamefont
  {Potier-Ferry}},\ }\href {\doibase 10.1016/j.jmps.2016.04.025} {\bibfield
  {journal} {\bibinfo  {journal} {J. Mech. Phys. Solids}\ }\textbf {\bibinfo
  {volume} {94}},\ \bibinfo {pages} {68} (\bibinfo {year} {2016})}\BibitemShut
  {NoStop}%
\bibitem [{\citenamefont {Xu}\ \emph {et~al.}(2017)\citenamefont {Xu},
  \citenamefont {Abdelmoula},\ and\ \citenamefont {Potier-Ferry}}]{Xu2017}%
  \BibitemOpen
  \bibfield  {author} {\bibinfo {author} {\bibfnamefont {F.}~\bibnamefont
  {Xu}}, \bibinfo {author} {\bibfnamefont {R.}~\bibnamefont {Abdelmoula}}, \
  and\ \bibinfo {author} {\bibfnamefont {M.}~\bibnamefont {Potier-Ferry}},\
  }\href {\doibase 10.1016/j.ijsolstr.2017.07.024} {\bibfield  {journal}
  {\bibinfo  {journal} {Int. J. Solids Struct.}\ }\textbf {\bibinfo {volume}
  {126--127}},\ \bibinfo {pages} {17} (\bibinfo {year} {2017})}\BibitemShut
  {NoStop}%
\bibitem [{\citenamefont {Jin}\ \emph {et~al.}(2015)\citenamefont {Jin},
  \citenamefont {Takei},\ and\ \citenamefont {Hutchinson}}]{Jin2015}%
  \BibitemOpen
  \bibfield  {author} {\bibinfo {author} {\bibfnamefont {L.}~\bibnamefont
  {Jin}}, \bibinfo {author} {\bibfnamefont {A.}~\bibnamefont {Takei}}, \ and\
  \bibinfo {author} {\bibfnamefont {J.~W.}\ \bibnamefont {Hutchinson}},\ }\href
  {\doibase 10.1016/j.jmps.2015.04.016} {\bibfield  {journal} {\bibinfo
  {journal} {J. Mech. Phys. Solids}\ }\textbf {\bibinfo {volume} {81}},\
  \bibinfo {pages} {22} (\bibinfo {year} {2015})}\BibitemShut {NoStop}%
\bibitem [{\citenamefont {Li}\ \emph {et~al.}(2011)\citenamefont {Li},
  \citenamefont {Jia}, \citenamefont {Cao}, \citenamefont {Feng},\ and\
  \citenamefont {Gao}}]{Li2011}%
  \BibitemOpen
  \bibfield  {author} {\bibinfo {author} {\bibfnamefont {B.}~\bibnamefont
  {Li}}, \bibinfo {author} {\bibfnamefont {F.}~\bibnamefont {Jia}}, \bibinfo
  {author} {\bibfnamefont {Y.-P.}\ \bibnamefont {Cao}}, \bibinfo {author}
  {\bibfnamefont {X.-Q.}\ \bibnamefont {Feng}}, \ and\ \bibinfo {author}
  {\bibfnamefont {H.}~\bibnamefont {Gao}},\ }\href {\doibase
  10.1103/physrevlett.106.234301} {\bibfield  {journal} {\bibinfo  {journal}
  {Phys. Rev. Lett.}\ }\textbf {\bibinfo {volume} {106}},\ \bibinfo {pages}
  {234301} (\bibinfo {year} {2011})}\BibitemShut {NoStop}%
\bibitem [{\citenamefont {Nabarro}(1982)}]{Nabarro1982}%
  \BibitemOpen
  \bibfield  {author} {\bibinfo {author} {\bibfnamefont {F.~R.~N.}\
  \bibnamefont {Nabarro}},\ }\href@noop {} {\emph {\bibinfo {title} {Theory of
  Dislocation}}}\ (\bibinfo  {publisher} {Dover, New York},\ \bibinfo {year}
  {1982})\BibitemShut {NoStop}%
\bibitem [{\citenamefont {Karampour}\ \emph {et~al.}(2013)\citenamefont
  {Karampour}, \citenamefont {Albermani},\ and\ \citenamefont
  {Gross}}]{Karampour2013}%
  \BibitemOpen
  \bibfield  {author} {\bibinfo {author} {\bibfnamefont {H.}~\bibnamefont
  {Karampour}}, \bibinfo {author} {\bibfnamefont {F.}~\bibnamefont
  {Albermani}}, \ and\ \bibinfo {author} {\bibfnamefont {J.}~\bibnamefont
  {Gross}},\ }\href {\doibase 10.1016/j.engstruct.2013.02.037} {\bibfield
  {journal} {\bibinfo  {journal} {Eng. Struct.}\ }\textbf {\bibinfo {volume}
  {52}},\ \bibinfo {pages} {317 } (\bibinfo {year} {2013})}\BibitemShut
  {NoStop}%
\bibitem [{\citenamefont {Vella}\ \emph {et~al.}(2009)\citenamefont {Vella},
  \citenamefont {Boudaoud},\ and\ \citenamefont {Adda-Bedia}}]{Vella2009}%
  \BibitemOpen
  \bibfield  {author} {\bibinfo {author} {\bibfnamefont {D.}~\bibnamefont
  {Vella}}, \bibinfo {author} {\bibfnamefont {A.}~\bibnamefont {Boudaoud}}, \
  and\ \bibinfo {author} {\bibfnamefont {M.}~\bibnamefont {Adda-Bedia}},\
  }\href {\doibase 10.1103/physrevlett.103.174301} {\bibfield  {journal}
  {\bibinfo  {journal} {Phys. Rev. Lett.}\ }\textbf {\bibinfo {volume} {103}},\
  \bibinfo {pages} {174301} (\bibinfo {year} {2009})}\BibitemShut {NoStop}%
\bibitem [{\citenamefont {Kolinski}\ \emph {et~al.}(2009)\citenamefont
  {Kolinski}, \citenamefont {Aussillous},\ and\ \citenamefont
  {Mahadevan}}]{Kolinski2009}%
  \BibitemOpen
  \bibfield  {author} {\bibinfo {author} {\bibfnamefont {J.~M.}\ \bibnamefont
  {Kolinski}}, \bibinfo {author} {\bibfnamefont {P.}~\bibnamefont
  {Aussillous}}, \ and\ \bibinfo {author} {\bibfnamefont {L.}~\bibnamefont
  {Mahadevan}},\ }\href {\doibase 10.1103/physrevlett.103.174302} {\bibfield
  {journal} {\bibinfo  {journal} {Phys. Rev. Lett.}\ }\textbf {\bibinfo
  {volume} {103}},\ \bibinfo {pages} {174302} (\bibinfo {year}
  {2009})}\BibitemShut {NoStop}%
\bibitem [{\citenamefont {Stoop}\ and\ \citenamefont
  {M\"uller}(2015)}]{Stoop2015}%
  \BibitemOpen
  \bibfield  {author} {\bibinfo {author} {\bibfnamefont {N.}~\bibnamefont
  {Stoop}}\ and\ \bibinfo {author} {\bibfnamefont {M.~M.}\ \bibnamefont
  {M\"uller}},\ }\href {\doibase 10.1016/j.ijnonlinmec.2015.02.013} {\bibfield
  {journal} {\bibinfo  {journal} {Int. J. Nonlinear Mech.}\ }\textbf {\bibinfo
  {volume} {75}},\ \bibinfo {pages} {115} (\bibinfo {year} {2015})}\BibitemShut
  {NoStop}%
\bibitem [{\citenamefont {Zhu}\ and\ \citenamefont {Chen}(2013)}]{Zhu2013}%
  \BibitemOpen
  \bibfield  {author} {\bibinfo {author} {\bibfnamefont {L.}~\bibnamefont
  {Zhu}}\ and\ \bibinfo {author} {\bibfnamefont {X.}~\bibnamefont {Chen}},\
  }\href {\doibase 10.1016/j.actbio.2013.04.011} {\bibfield  {journal}
  {\bibinfo  {journal} {Acta Biomater.}\ }\textbf {\bibinfo {volume} {9}},\
  \bibinfo {pages} {7968 } (\bibinfo {year} {2013})}\BibitemShut {NoStop}%
\bibitem [{Sup()}]{Sup}%
  \BibitemOpen
  \href@noop {} {\bibinfo  {journal} {See Supplemental Material in the end for
  details on theoretical derivations, numerical simulations, discussions on
  curvature effects and comparisons of constitutive laws}\ }\BibitemShut
  {NoStop}%
\bibitem [{\citenamefont {Koiter}(1945)}]{Koiter1945}%
  \BibitemOpen
\bibfield  {journal} {  }\bibfield  {author} {\bibinfo {author} {\bibfnamefont
  {W.}~\bibnamefont {Koiter}},\ }\emph {\bibinfo {title} {On the Stability of
  Elastic Equilibrium}},\ \href@noop {} {Ph.D. thesis},\ \bibinfo  {school}
  {Delft University of Technology, Netherlands} (\bibinfo {year}
  {1945})\BibitemShut {NoStop}%
\bibitem [{\citenamefont {Yamaki}(1984)}]{Yamaki1984}%
  \BibitemOpen
  \bibfield  {author} {\bibinfo {author} {\bibfnamefont {N.}~\bibnamefont
  {Yamaki}},\ }\href@noop {} {\emph {\bibinfo {title} {Elastic Stability of
  Circular Cylindrical Shells}}}\ (\bibinfo  {publisher} {North-Holland,
  Amsterdam},\ \bibinfo {year} {1984})\BibitemShut {NoStop}%
\bibitem [{\citenamefont {van~der Heijden}(2009)}]{Koiter2009}%
  \BibitemOpen
  \bibfield  {author} {\bibinfo {author} {\bibfnamefont {A.~M.~A.}\
  \bibnamefont {van~der Heijden}},\ }\href@noop {} {\emph {\bibinfo {title}
  {W.T. Koiter's Elastic Stability of Solids and Structures}}}\ (\bibinfo
  {publisher} {Cambridge University Press, New York},\ \bibinfo {year}
  {2009})\BibitemShut {NoStop}%
\bibitem [{\citenamefont {Holzapfel}(2000)}]{Holzapfel2000}%
  \BibitemOpen
  \bibfield  {author} {\bibinfo {author} {\bibfnamefont {G.~A.}\ \bibnamefont
  {Holzapfel}},\ }\href@noop {} {\emph {\bibinfo {title} {Nonlinear Solid
  Mechanics: A Continuum Approach for Engineering}}}\ (\bibinfo  {publisher}
  {Wiley, Chichester, England},\ \bibinfo {year} {2000})\BibitemShut {NoStop}%
\bibitem [{\citenamefont {Hunt}\ \emph {et~al.}(2000)\citenamefont {Hunt},
  \citenamefont {Peletier}, \citenamefont {Champneys}, \citenamefont {Woods},
  \citenamefont {{Ahmer Wadee}}, \citenamefont {Budd},\ and\ \citenamefont
  {Lord}}]{Hunt2000}%
  \BibitemOpen
  \bibfield  {author} {\bibinfo {author} {\bibfnamefont {G.~W.}\ \bibnamefont
  {Hunt}}, \bibinfo {author} {\bibfnamefont {M.~A.}\ \bibnamefont {Peletier}},
  \bibinfo {author} {\bibfnamefont {A.~R.}\ \bibnamefont {Champneys}}, \bibinfo
  {author} {\bibfnamefont {P.~D.}\ \bibnamefont {Woods}}, \bibinfo {author}
  {\bibfnamefont {M.}~\bibnamefont {{Ahmer Wadee}}}, \bibinfo {author}
  {\bibfnamefont {C.~J.}\ \bibnamefont {Budd}}, \ and\ \bibinfo {author}
  {\bibfnamefont {G.~J.}\ \bibnamefont {Lord}},\ }\href {\doibase
  10.1023/A:1008398006403} {\bibfield  {journal} {\bibinfo  {journal}
  {Nonlinear Dynam.}\ }\textbf {\bibinfo {volume} {21}},\ \bibinfo {pages} {3}
  (\bibinfo {year} {2000})}\BibitemShut {NoStop}%
\bibitem [{\citenamefont {Almroth}\ \emph {et~al.}(1964)\citenamefont
  {Almroth}, \citenamefont {Holmes},\ and\ \citenamefont
  {Brush}}]{Almroth1964}%
  \BibitemOpen
  \bibfield  {author} {\bibinfo {author} {\bibfnamefont {B.~O.}\ \bibnamefont
  {Almroth}}, \bibinfo {author} {\bibfnamefont {A.~M.~C.}\ \bibnamefont
  {Holmes}}, \ and\ \bibinfo {author} {\bibfnamefont {D.~O.}\ \bibnamefont
  {Brush}},\ }\href {\doibase 10.1007/BF02323088} {\bibfield  {journal}
  {\bibinfo  {journal} {Exp. Mech.}\ }\textbf {\bibinfo {volume} {4}},\
  \bibinfo {pages} {263} (\bibinfo {year} {1964})}\BibitemShut {NoStop}%
\bibitem [{\citenamefont {Horton}\ and\ \citenamefont
  {Durham}(1965)}]{Horton1965}%
  \BibitemOpen
  \bibfield  {author} {\bibinfo {author} {\bibfnamefont {W.~H.}\ \bibnamefont
  {Horton}}\ and\ \bibinfo {author} {\bibfnamefont {S.~C.}\ \bibnamefont
  {Durham}},\ }\href {\doibase 10.1016/0020-7683(65)90015-6} {\bibfield
  {journal} {\bibinfo  {journal} {Int. J. Solids Struct.}\ }\textbf {\bibinfo
  {volume} {1}},\ \bibinfo {pages} {59} (\bibinfo {year} {1965})}\BibitemShut
  {NoStop}%
\end{thebibliography}%

\end{document}